\documentclass[11pt,a4paper]{article}

\usepackage{hyperref,graphicx,color,amsmath,amssymb,slashed,mcite,xspace}

\newcommand{\pythia}{\textsc{Pythia}\xspace}
\newcommand{\madgraph}{\textsc{MadGraph5}\xspace}
\newcommand{\madgraphamc}{\textsc{MadGraph5\_aMC@NLO}\xspace}
\newcommand{\delphes}{\textsc{Delphes}\xspace}

\begin{document}

\begin{titlepage}
\begin{flushright}
UT--14--24\\
IPMU14--0116
\end{flushright}
\vskip 3cm
\begin{center}
{\Large \bf
Reconstruction of vectorlike top partner\\ 
from fully hadronic final states
}
\vskip 1.5cm
{
Motoi Endo$^{(a,b)}$,
Koichi Hamaguchi$^{(a,b)}$,
Kazuya Ishikawa$^{(a)}$,
and
Martin Stoll$^{(a)}$
}
\vskip 0.9cm
{\it $^{(a)}$ Department of Physics, University of Tokyo, Bunkyo-ku, Tokyo 113--0033, Japan \vspace{0.2cm}
\par
$^{(b)}$ Kavli Institute for the Physics and Mathematics of the Universe (Kavli IPMU), \\
University of Tokyo, Kashiwa 277--8583, Japan
}
\vskip 2cm
\abstract{
We investigate the potential to search for the vectorlike top partner
in fully hadronic final states at the LHC.
An algorithm is developed that kinematically reconstructs the  vectorlike top.
 We show that for moderate masses and a large branching fraction
into the top quark and Higgs boson, 
the reconstruction works with good quality. 
}
\end{center}
\end{titlepage}

\section{Introduction}

The discovery of the Higgs boson in 2012 at the LHC~\cite{Aad:2012tfa,Chatrchyan:2012ufa} completed the particle content of the Standard Model (SM) and triggered a new era of physics beyond the SM.
As the LHC will restart soon in 2015 at $\sqrt{s}=13$--$14$~TeV, it is important to explore a variety of possible scenarios that can be probed at this new energy frontier.
In this paper, we discuss the possibility of searching for a vectorlike top partner and propose a new approach to reconstruct it from its decay into fully hadronic final states.

The vectorlike top partner is a heavy quark that has electric charge $2/3$.
It is typically assumed to mainly couple to the third-generation quarks of the SM.
In supersymmetric (SUSY) models, such vectorlike matters can increase the light Higgs boson mass while keeping other SUSY particles relatively light~\cite{Moroi:1992zk,*Babu:2004xg,*Babu:2008ge,*Martin:2009bg,Endo:2011mc,*Moroi:2011aa,*Endo:2011xq,*Endo:2012cc,Asano:2011zt,*Evans:2011uq,*Martin:2012dg,*Fischler:2013tva}.
This is one of the few viable SUSY models that can explain the 126 GeV Higgs boson mass and the discrepancy of the muon anomalous magnetic moment simultaneously~\cite{Endo:2011mc}.
Another class of well-motivated models with vectorlike tops are little Higgs models~\cite{ArkaniHamed:2001nc,*ArkaniHamed:2002pa,*ArkaniHamed:2002qx,*Low:2002ws,*Kaplan:2003uc}, where the top partner is introduced to cut off otherwise quadratically divergent loop integrals. 

In all these models, the vectorlike top is expected to be directly produced at the LHC. 
Searches for pair-production of vectorlike tops ($t'$) have been conducted for several final states available from the $t'\rightarrow t h$,  $t'\rightarrow t Z$ and $t'\rightarrow b W$ decay channels. 
Current exclusion bounds on the vectorlike top mass at $\sqrt{s}=8$~TeV  are about 690--780 GeV from CMS~\cite{Chatrchyan:2013uxa} and 550--850 GeV from ATLAS~\cite{ATLAS-CONF-2013-018,*ATLAS-CONF-2013-051,*ATLAS-CONF-2013-056,*ATLAS-CONF-2013-060}, depending on the assumed branching ratios. 
In these studies, a subsequent (semi)leptonic decay is used as a typical search channel. 

In this paper, we investigate the possibility of searching for a vectorlike top partner from purely hadronic final states at the LHC, assuming that the vectorlike tops are pair-produced and dominantly decay into $t$ and $h$.\footnote{See Refs.~\cite{Kribs:2010ii,Azatov:2012rj,Harigaya:2012ir,Girdhar:2012vn,Okada:2012gy,Gopalakrishna:2013hua,Bhattacharya:2013iea,Girdhar:2014wua} for previous studies
on $t'\to th$ from pair production.}
For a heavy vectorlike top, its decay products are considerably boosted and hence subsequent decay products of each $t$ and $h$ are collimated in one area of the detector.
We apply substructure methods~\cite{Butterworth:2008iy,Plehn:2010st} to identify the top quark and the Higgs boson within these large ``fat'' jets.
We also propose an algorithm to determine the $t$-$h$ combination based on a massive pair hypothesis.
We show that for moderate masses of the vectorlike top, decent event rates are feasible within the first period of the LHC run II and find that the vectorlike top can be reconstructed with good quality.
 
This paper is organized as follows. 
After briefly introducing the model in Sec.~\ref{sec:model}, we describe the setup of our simulation in Sec.~\ref{sec:Event_Generation}.
A detailed description of cuts and algorithms is given in Sec.~\ref{sec:analysis}.
The main results of our simulation are summarized in Sec.~\ref{sec:results}.
We conclude our findings and give a brief outlook in Sec.~\ref{sec:summary_outlook}.
  
\section{Model}
\label{sec:model}

In this paper we consider the decay of the vectorlike top ($t'$) into top ($t$) and Higgs ($h$), which is described by the following Lagrangian\footnote{In general, there is also a model-dependent term $\lambda h \bar{t}\gamma_5 t' + h.c. $ in the Lagrangian which can give the top quark a dominant chirality. We however expect that our results do not change significantly in the presence of this term because our algorithm is blind with respect to the chirality of the top quark, although a detailed study would be necessary to quantify the effect. In our analysis, we assume $\lambda=0$ for simplicity.}
\begin{align}
 \mathcal{L} = \mathcal{L}_\text{SM} + \bar{t'} \left( \imath\slashed{D} - m_{t'} \right) t'
      + y_{t'} h \bar{t} t' + {\it h.c.} \,.
\end{align}
 
We investigate pair production of vectorlike tops at the LHC with a center-of-mass energy of $14\text{ TeV}$,
\begin{align}
 pp\to t'\bar{t'}\,,
\end{align}
and consider the following decay chain to fully hadronic final states,
\begin{align}
 t'\to th\to bjj\; b\bar{b}\,,
\end{align}
where $j$ denotes $u$,$d$,$c$ or $s$ (anti)quarks.
For simplicity, we assume that the vectorlike top decays exclusively to top and Higgs. 
As for the mass of the $t'$, we consider $m_{t'}=800$ and $900\text{ GeV}$.
The top quark mass is taken to be $173.5\text{ GeV}$ and we assume the SM Higgs boson branching ratio $\text{BR}(h\rightarrow b\bar{b})=0.56$ with a Higgs mass of $126\text{ GeV}$~\cite{Beringer:1900zz}.
 
\section{Event generation}
\label{sec:Event_Generation}

All events are simulated with \madgraph~1.5.14~\cite{Alwall:2011uj} in combination with \pythia~6.4~\cite{Sjostrand:2006za} and the \delphes~3 fast detector simulation~\cite{deFavereau:2013fsa}.
The parameters of the latter are adjusted to the ATLAS detector as provided by the \madgraph package.\footnote{Parameters for jet clustering and bottom tagging will be discussed later.}
Common cuts are imposed on all final-state partons at generator level: transverse momentum $p_T\geq20\text{ GeV}$ and mutual separation $\Delta R\equiv \sqrt{\Delta\phi^2 + \Delta\eta^2} \geq 0.4$, where $\phi$ and $\eta$ are the parton's azimuthal angle and its pseudorapidity.

The main background processes are $bb\bar{b}\bar{b}$, $t\bar{t}$, $t\bar{t}b\bar{b}$, and $t\bar{t}h$ after imposing all cuts described in the next section.
Other processes like $b\bar{b}V$, $b\bar{b}h$, $t\bar{b} + \bar{t}b$, $b\bar{b}$, and $t\bar{t}V$ turned out negligible. 
Pure multijet QCD background events are difficult to simulate reliably, but we expect that they are also efficiently suppressed by our cut procedure, in particular by multiple $b$-tagging.
  
Both for signal and backgrounds, we generate events at leading order (LO) and rescale them by uniform K factors assuming the event distribution is not affected much at next-to-leading order (NLO).
For signal events, the cross section is calculated at NLO using \madgraphamc~\cite{Alwall:2011uj}.
We obtain the K factors $1.33$ for $m_{t'}=800\text{ GeV}$ and $1.32$ for $m_{t'}=900\text{ GeV}$. 

For background processes, limited computational power forces us to impose additional severe generator-level cuts.
We thus demand large generator-level scalar transverse momentum, $H_T^\text{p.l.} \equiv \sum_{\{\text{partons } i\}} p_T^{(i)} \geq 1000\text{ GeV}$. 
In this way a larger fraction of generated events can be obtained in the signal region.
Note that signal events tend to have large $H_T$ and a more severe cut will be imposed at detector level, cf.~Sec.~\ref{sec:analysis:ht}.
On the other hand, this parton level cut cannot be efficiently applied to event generation at NLO, because it acts differently on events with different final-state multiplicity (the set of partons which contribute to the sum is different).\footnote{For the same reason approximate methods such as MLM matching~\cite{Mangano:2006rw} are also not feasible.}
Thus, we generate background events at LO.
 
We are interested in the background cross sections only after a cut on $H_T$ is imposed at detector level.
The values at LO can be obtained by cutting on generated events.\footnote{To achieve better accuracy, these events are produced with a lower cut, $H_T^\text{p.l.}\geq600\text{ GeV}$.}
Results are then rescaled by uniform K-factors which we take as $1.40$ for $bb\bar{b}\bar{b}$~\cite{Worek:2013zwa}, $1.61$ for $t\bar{t}$~\cite{Kidonakis:2008mu}, $1.77$ for $t\bar{t}b\bar{b}$~\cite{Bevilacqua:2009zn}, and $1.10$ for $t\bar{t}h$~\cite{Frederix:2011zi}.
We do not attempt to estimate uncertainties of the background cross sections, as these values should be measured experimentally from appropriate control regions.
Consequently, this paper does not show a cut-and-count analysis but rather demonstrates the potential of reconstructing the vectorlike top.

\section{Analysis}
 \label{sec:analysis}

This paper aims at developing an analysis that can kinematically reconstruct the vectorlike top particle.
First, general cuts reflecting the high-energy deposit and multiple-$b$ nature of the signal are employed, which enhances the signal-to-background ratio.
In a next step, top quark and Higgs boson candidates are reconstructed.
Finally the four-momentum of the vectorlike top is recovered which gives access to the reconstructed mass.
 
We propose the following analysis:
\begin{itemize}
 \item large-$H_T$ cut
 \item multiple bottom cut
 \item top tagging and cut
 \item Higgs tagging and cut
 \item vectorlike top reconstruction
\end{itemize}
Each of the keywords listed here is further explained in a dedicated subsection.
See also Table~\ref{table:signal_region} for an overview of the signal regions.
  
We consider that the large $H_T$ cut also serves to trigger events. 
For the case that this will not be adopted at the 14 TeV LHC
we investigated the following event triggers as well: 
4 jets each with $p_T\ge 90$~GeV  or 
5 jets each with $p_T\ge 55$~GeV
(cf.~Refs.~\cite{Aad:2009wy,Trocino:1623353,Guiducci:1561429}).
It was found that the final results of our analysis do not change under these additional cuts.

\begin{table}[t]
 \begin{center}
  \begin{tabular}{|c||c|c|c|}\hline
   & \hspace{0.5cm}SR1\hspace{0.5cm} &\hspace{0.5cm} SR2\hspace{0.5cm} &\hspace{0.5cm} SR3\hspace{0.5cm} \\ \hline \hline
   $H_T$ & \multicolumn{3}{|c|}{$\geq$ 1200~GeV}\\ \hline
   number of tagged $b$ & \multicolumn{3}{|c|}{$\geq$ 4}\\ \hline
   number of tagged $t$ & = 1 & = 2 & = 2\\ \hline
   number of tagged $h$ & = 2 & = 1 & = 2\\ \hline
  \end{tabular}
  \caption{The signal regions. For SR1 and SR2, we demand additional conditions for reconstructing vectorlike tops (see Sec.~\ref{sec:analysis:massive_pair}).}
  \label{table:signal_region}
 \end{center}
\end{table}
 
\subsection{Scalar transverse momentum cut}
\label{sec:analysis:ht}
 
\begin{figure}[tb]
 \begin{center}
  \includegraphics[scale=.5]{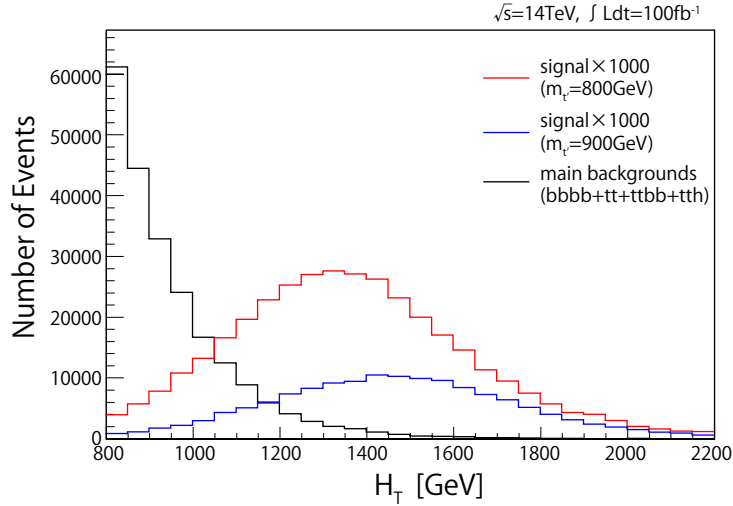}
  \caption{$H_T$ distribution for signal and background processes (detector level). The red (blue) line corresponds to signal events with $m_{t'}=800\ (900)\text{ GeV}$ and is 1000 times enlarged, and the black line describes the main backgrounds which contain $bb\bar{b}\bar{b}$, $t\bar{t}$, $t\bar{t}b\bar{b}$ and $t\bar{t}h$ processes.}
  \label{fig:cut_ht_dist}
 \end{center}
\end{figure}

In order to suppress continuum backgrounds we impose a cut on scalar transverse momentum, given by
\begin{align}
 H_T \equiv \sum_{\text{jets }j} p_T^{(j)} \,.\label{eq:ht_def}
\end{align}
Here and for $b$ tagging we use the anti-$k_T$ algorithm~\cite{Cacciari:2008gp} as implemented in FastJet~\cite{Cacciari:2011ma} with parameters $R=0.4$ and $p_T \geq 20\text{ GeV}$ for jet clustering.
The heavy vectorlike top's decay exhibits a typically large value of order $H_T \sim {\cal O}(2 m_{t'})$ whereas the cross sections of all standard model processes drop exponentially.
$H_T$ distributions of signal and background events are shown in Fig.~\ref{fig:cut_ht_dist}.
We therefore require
\begin{align}
 H_T \geq 1200\text{ GeV} \,.
\end{align}
 
\subsection{Bottom tagging and cut}
 
As the signal contains six bottom quarks in the final state, a cut on the number of $b$-tagged jets is indicated. 
$b$ tagging is performed with an algorithm identical to the default in \delphes~\cite{deFavereau:2013fsa}.
We choose a working point where $b$-initiated jets are correctly identified with $70\%$ probability, $\epsilon_\text{tag}=0.70$, and assume the fractions of jets which are misidentified as bottom quark-initiated to be $\epsilon_\text{mis}^{(udsg)}=0.01$ for light jets (light quark- and gluon-initiated jets)  and $\epsilon_\text{mis}^{(c)}=0.10$ for charm-initiated jets.\footnote{The tagging efficiencies quoted by ATLAS Collaboration are $\epsilon_\text{mis}^{(udsg)}\simeq0.01$, $\epsilon_\text{mis}^{(c)}\simeq0.20$ for $\epsilon_\text{tag}=0.70$ at $\sqrt{s}=7\text{ TeV}$~\cite{ATLAS:2012ima} and are expected to be improved at the 14 TeV LHC.}

\begin{figure}[tb]
 \begin{center}
  \includegraphics[scale=.45]{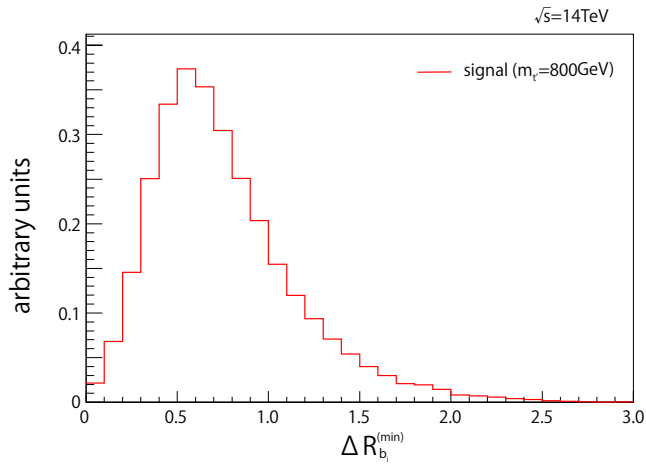}
  \caption{Distribution of bottom quark isolation for signal events (parton level). The horizontal axis corresponds to the smallest distance between each (anti)bottom quark and any other particle in the partonic final state, $\Delta R_{b_i}^{\text{(min)}}=\min_{j\ne b_i} \Delta R(b_i,j)\quad (i=1,\cdots,6)$.}
  \label{fig:cut_bottom_jetseparation}
 \end{center}
\end{figure}

The tagging  efficiencies may not be applicable if there is overlap between bottom-initiated and other jets.
In Fig.~\ref{fig:cut_bottom_jetseparation}, the distribution of the minimal distance $\Delta R=\sqrt{(\Delta \phi)^2+(\Delta \eta)^2}$ between each (anti)bottom quark $b_i$ ($i=1,\cdots,6$) and any other particle in the partonic final state is shown for signal events,
\begin{align}
 \Delta R_{b_i}^{\text{(min)}}
 =
 \min_{j\ne b_i} \Delta R(b_i,j)\,,
\end{align}
where $j$ runs over all partons (including $b$).
Here, the generator level cut $\Delta R\ge 0.4$ is not imposed.
As can be seen from the figure, the vast majority of $b$-quarks are separated from any other parton by a distance greater than the jet clustering radius $\Delta R=0.4$.

We require at least 4 $b$-tagged jets in this analysis, which is sufficient for an effective rejection of SM background events while retaining reasonable signal event rates.
  
\subsection{Fat jets}

For the mass of the vectorlike top $m_{t'}\geq 800\text{ GeV}$ considered in this paper, its decay products $t$ and $h$ are typically boosted, with $p_T^{t,h}\gtrsim 200\text{ GeV}$.
The final state jets emerging from the subsequent decay $t\rightarrow b j j$ (and $h\rightarrow b\bar{b}$ respectively) will therefore be collimated with a typical distance $\Delta R_\text{daughters} \sim 2m_\text{mother} / p_T$ and can be caught within a fat jet of large radius.
For boosted top quarks, the HEPTopTagger~\cite{Plehn:2010st} proved very successful in this kinematic regime by looking at the substructure of a fat jet with radius $\Delta R=1.5$ and $p_T^\text{fat jet}\geq200\text{ GeV}$.
Due to the high-multiplicity final state, a tagger based on jet substructure should be preferred over a combinatoric algorithm.
 
In this paper, we refer to fat jets as jets which are clustered from calorimeter information using the Cambridge-Aachen algorithm~\cite{Dokshitzer:1997in,Wobisch:1998wt} with parameters $\Delta R=1.5$ and $p_T^\text{fat jet}\geq200\text{ GeV}$.
We treat fat jets emerging from $t$ or $h$ on equal footage.
 
\begin{figure}[t]
\begin{minipage}{0.48\hsize}
  \begin{center}
    \includegraphics[scale=.38]{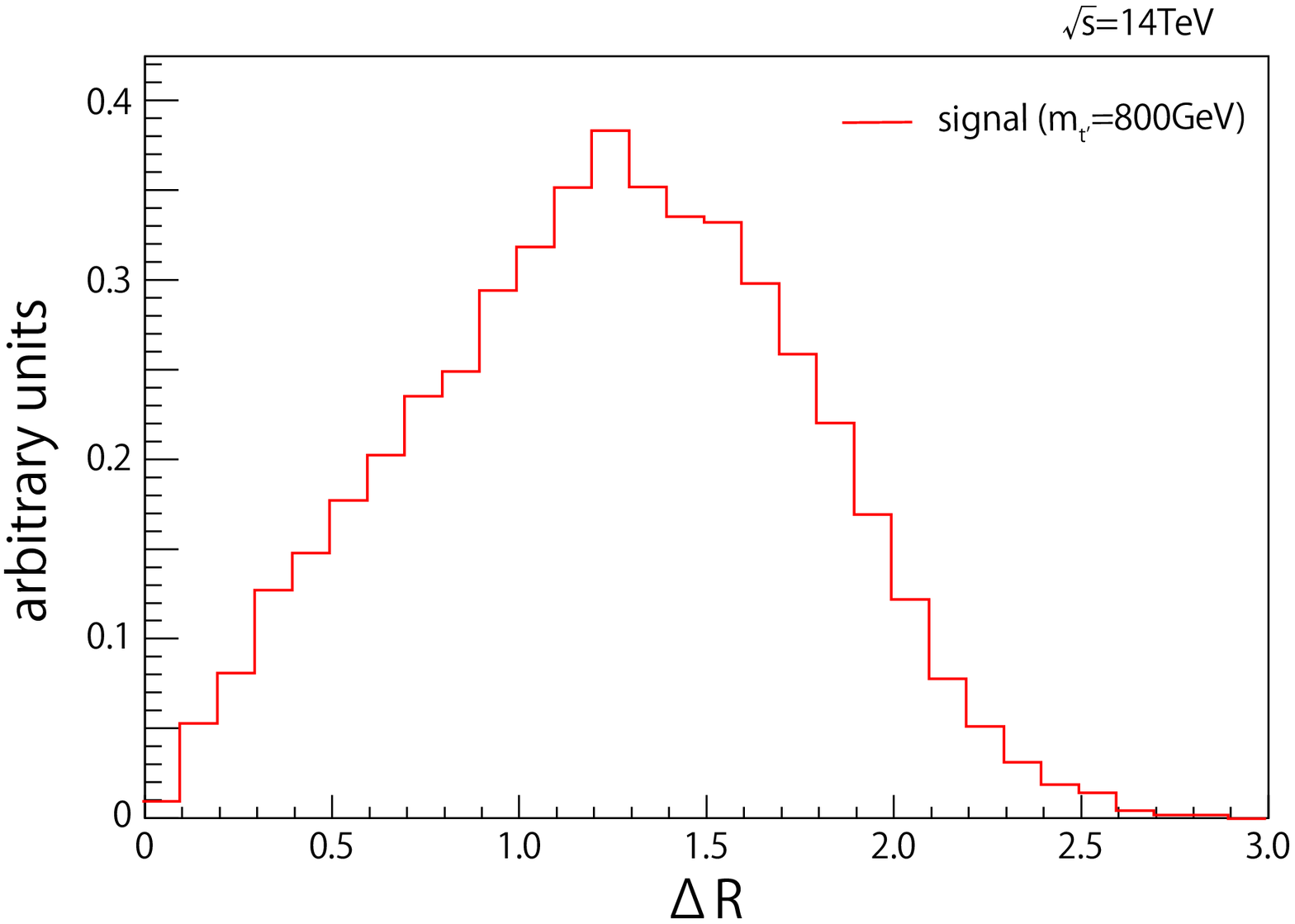}
  \end{center}
\end{minipage}
\begin{minipage}{0.48\hsize}
  \begin{center}
    \includegraphics[scale=.38]{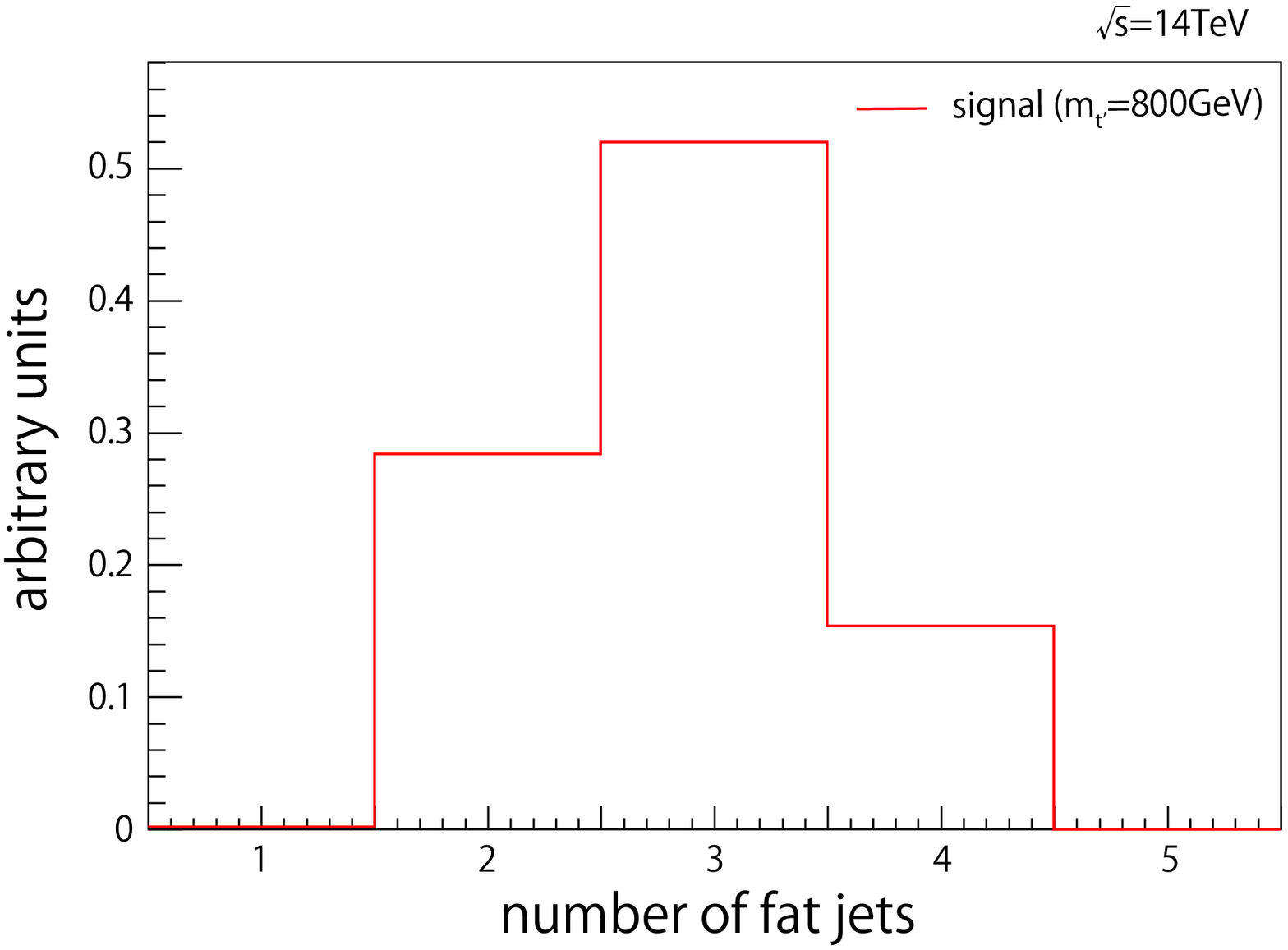}
  \end{center}
\end{minipage}
    \caption{(left) The smallest distance between any two of the top quarks and Higgs bosons
 in $pp\rightarrow t' \bar{t'}\rightarrow t h \bar{t} h$ (parton level).
    (right) Distribution of the number of reconstructed fat jets for the signal (detector level).}
    \label{fig:cut_fat_figs}
\end{figure}

Fig.~\ref{fig:cut_fat_figs} (lhs) shows the distribution of the smallest distance between any two of the top quarks and Higgs bosons in $pp\rightarrow t' \bar{t'}\rightarrow t h \bar{t} h$.
It is generally smaller than the fat jet radius $\Delta R=1.5$ and thus in a typical event (at least) one fat jet  contains the decay products of two partons. 
In most signal events the number of reconstructed fat jets is indeed less than four, see the right-hand side plot.
 
At least three fat jets are required as candidates for top and Higgs in SR1 and SR2, and at least four in SR3.
 
\subsection{Top quark tagging and reconstruction}

We rely on the HEPTopTagger~\cite{Plehn:2010st} to tag and kinematically reconstruct boosted tops.
As the concept is very similar to the our Higgs tagger implementation (see next subsection), we briefly go over the algorithm.
The following procedure is applied to each fat jet. 
\begin{enumerate}
 \item First, the fat jet is successively declustered with a mass-drop criterion. At each step in the iterative un-doing of the last clustering of the jet $j$, both subjets $j_1,j_2$ are kept only if a substantial mass drop occurs (corresponding to the two-body decay of a heavy particle). Otherwise the less massive subjet is removed. The mass-drop condition reads $\max m_{j_i}<0.8 m_j$. Also, subjets with $m_j<30\text{ GeV}$ are not further decomposed, which eventually ends the un-clustering stage.

 \item Additional soft radiation is then removed by applying a filtering stage and the five hardest subjets are kept.
  
 \item A top quark candidate is reconstructed from three subjets if the combined mass is within $150\leq m_{jjj} \leq 200$~GeV and various subjet mass ratios resemble a real top decay.\footnote{These cuts are chosen to effectively reject background events. See Ref.~\cite{Plehn:2010st} for a detailed discussion. We adopt all parameters as described therein.}
  
 \item We require these three subjets to mutually meet the condition $\Delta R \geq 0.4$ to be consistent with a similar cut at event generation level.
  
 \item If there are multiple top candidates, the one with a mass closest to the real top quark mass is chosen.
\end{enumerate}
The conditions 4 and 5 are different from the original HEPTopTagger~\cite{Plehn:2010st}.
In particular, in the original paper a tag was realised if and only if the subjet combination with a combined mass closest to the top quark mass passes all cuts.
Due to this modification, signal and background mistag rates are  similarly enhanced in our analysis.

If a top candidate is reconstructed, the corresponding fat jet is not considered as Higgs candidate.
In our analysis, we require 1 or 2 tagged tops in a given event.

\subsection{Higgs boson tagging and reconstruction}

Higgs tagging proceeds very similarly to top tagging.
We implemented an algorithm loosely based on the BDRS Higgs tagger~\cite{Butterworth:2008iy}.
A good review of various tagging algorithms can be found in Ref.~\cite{Plehn:2011tg}.
 
\begin{enumerate}
 \item The unclustering stage is similar to the HEPTopTagger described above. However in addition to the mass-drop criterion (which in this case reads $\max m_{j_i}<0.67 m_j$), a symmetry requirement is imposed: $\min p_{T,j_i} / \max p_{T,j_i} > 0.09$ which reflects the splitting $h\rightarrow b\bar{b}$.\footnote{The parameters are the same as in the BDRS Higgs tagger~\cite{Butterworth:2008iy}.}
  
 \item The filtering stage is identical to the HEPTopTagger. Note that we also keep the five hardest subjets, although in $h$-induced fat jets keeping only the three hardest subjets (as suggested in Ref.~\cite{Butterworth:2008iy}) would yield better background discrimination. As a significant number of fat jets contain decay products of another $t$ or $h$, this choice allows efficient tagging of those contaminated fat jets as well.
  
 \item A Higgs boson candidate is reconstructed from two filtered subjets in the mass range $100\leq m_{jj}\leq150\text{ GeV}$. 
  
 \item We require all Higgs candidate subjets to mutually meet the condition $\Delta R \geq 0.4$ for consistency with event generation.
  
 \item If multiple candidates arise, the one with a mass closest to the real Higgs mass is selected.
\end{enumerate}
Unlike suggested in Ref.~\cite{Butterworth:2008iy}, we do not require a tagged $b$ jet inside the reconstructed Higgs.\footnote{We investigated this option and found improved purity and slightly better signal-to-background ratios, but at the cost of smaller signal event numbers. It should be considered once higher integrated luminosity is available.}

We examined our algorithm with clean samples of $pp\rightarrow Zh\rightarrow (\mu^+\mu^-)(b\bar{b})$ events where $p_T^h \simeq 250\text{ GeV}$.
The tagging efficiency, i.e.~the fraction of tagged events, turned out to be $50\!\sim\!60\%$.
Misidentification rates are strongly process dependent.
 
We demand 1 or 2 reconstructed Higgs bosons in this analysis.

\subsection{Massive pair hypothesis and reconstructed mass}
\label{sec:analysis:massive_pair}

The vectorlike top mass is kinematically reconstructed from the tagged top quark and Higgs boson momenta as 
\begin{align}
 M(t,h) = \sqrt{ (p_t^\mu+p_h^\mu)^2 } \,,
\end{align}
where $p_i^\mu$ is the four-momentum of given particle $i$.
However, if there are two tagged tops and/or two tagged Higgs bosons, it is not clear how to assign which top and Higgs have emerged from the same vectorlike top.
  
In the case that both two tops ($t_1,t_2$) and two Higgs bosons ($h_1,h_2$) are reconstructed in an event (SR3), there are two possible combinations for the vectorlike tops, $\{(t_1,h_1),(t_2,h_2)\}$ and $\{(t_1,h_2),(t_2,h_1)\}$.
In the true combination, the two reconstructed masses should be similar since we consider vectorlike top pair production.
We thus choose the combination which gives a smaller mass difference,
\begin{align}
 \min\left[ |M(t_1, h_1)-M(t_2,h_2)|, |M(t_1, h_2)-M(t_2,h_1)| \right] \,.
\end{align} 
   
Next, let us consider the case where one top ($t$) and two Higgs bosons ($h_1,h_2$) are reconstructed (SR1).
In this case, three out of four particle momenta are known, $p^\mu_t,p^\mu_{h_1},p^\mu_{h_2}$.
We suggest the following algorithm to determine the correct pairing.
Under the signal hypothesis, the momentum of the undetected fourth particle (denoted as $t_\text{miss}$) obeys the following constraints:
\begin{eqnarray}
 &&\vec{p}_{T,t_\text{miss}}+\sum_{i=t,h_1,h_2}\vec{p}_{T,i}=0\,,\\
 &&\left(p^\mu_{t_\text{miss}}\right)^2=m_t^2 \,.
\end{eqnarray}
The former equation is due to the absence of missing energy in the fully hadronic final states.
From these equations, the longitudinal momentum component $p_{z,t_\text{miss}}$ is the only unknown parameter.
To determine $p_{z,t_\text{miss}}$, we demand the two reconstructed vectorlike tops to have equal masses.
Because there are two possible combinations for $t$-$h$ pairs, we arrive at two equations,
\begin{eqnarray}
 M(t,h_1)&=&M(t_\text{miss},h_2)\label{eq:mass_1} \,,\\
 \text{or}\quad
 M(t,h_2)&=&M(t_\text{miss},h_1)\label{eq:mass_2} \,.
\end{eqnarray}
We choose the combination using the following criteria.
\begin{itemize}
 \item[(a)] In the case where there is no solution for either of the equations, the event is inconsistent with the signal hypothesis and is discarded.
 
 \item[(b)] If exactly one of the equations yields a solution, the combination of $t$ and $h$ is uniquely determined. The reconstructed mass of the vectorlike top then is $M(t,h_1)$ (Eq.~(\ref{eq:mass_1}) is solvable) or $M(t,h_2)$ (Eq.~(\ref{eq:mass_2}) is solvable).
 
\item[(c)] 
If both equations give solutions, we choose the one with minimal pseudorapidity $\eta_{t_\text{miss}}$ and the vectorlike top mass is reconstructed from the corresponding $t$-$h$ pair.\footnote{Note that there can be two solutions for each equation. In our algorithm, we try to avoid any bias on the reconstructed mass. Once the order of the vectorlike top mass is known, the selection criterion can be optimized accordingly.}
At parton level, this choice agrees with the Monte Carlo truth with roughly $2/3$ accuracy.

 \end{itemize}
Note that in any case we do not use the fourth particle's momentum $p^\mu_{t_\text{miss}}$ to reconstruct the vectorlike top.
 
The case where two tops and one Higgs boson are reconstructed (SR2) is analyzed analogously.

For the signal regions SR1 and SR2, about a few percent, 30\% and 70\% of signal events fall in categories of (a), (b) and (c) respectively.
 
\section{Results}
\label{sec:results}

\begin{table}[hbtp]
 \begin{center}
  \begin{tabular}{|c||c||c||c||c|c|c|c|} \hline
   Process & \multicolumn{2}{|c||}{$t'\bar{t'}$} &b.g. & $bb\bar{b}\bar{b}$ & $t\bar{t}$ & $t\bar{t}b\bar{b}$ & $t\bar{t}h$   \\ \hline
    & 800~GeV & 900~GeV & &  &  &  &    \\\hline\hline
   Cross section[fb] & 3.75 & 1.52 & --- & 2.20$\times10^6$ & 1.39$\times10^5$ & 494 & 25.5  \\ \hline
   \multicolumn{8}{l}{number of events for 100 fb$^{-1}$} \\ \hline
   $H_T\geq1200\text{ GeV}$ & 266 & 123 & 14800 & 5320 & 9120 & 373 & 29.6  \\\hline
   $\# b \geq 4$ & 185 & 84.6 & 1560 & 1240 & 210 & 100 & 8.5 \\\hline \hline
   SR1 & 25.0 & 11.3 & 10.3 & 2.7 & 2.9 & 4.2 & 0.5 \\\hline
  SR2 & 13.0 & 5.7 & 5.8 & 0.7 & 2.3 & 2.5 & 0.3 \\\hline
  SR1+SR2 & 38.0 & 17.0 & 16.1 & 3.4 & 5.2 & 6.7 & 0.8 \\\hline
 \end{tabular}
 \caption{Cross sections and event numbers for an integrated luminosity of 100~fb$^{-1}$ at the LHC with $\sqrt{s}=14\text{ TeV}$. Results for the signal are shown separately for two different masses of the vectorlike top, 800 and 900~GeV. The sum of all relevant background processes (``b.g.'') as well as their individual breakdown is given in the right-hand columns. For the definition of the signal regions (SR) see Table~\ref{table:signal_region}. In SR3, the number of signal events for 800~GeV turns out to be less than two with almost vanishing backgrounds $< 0.35$.}
 \label{tab:result_nofevents}
 \end{center}
\end{table}
 
Event numbers under the cuts described above are shown in Table~\ref{tab:result_nofevents} for signal regions SR1 and SR2.
All numbers are rescaled to an integrated luminosity of $100\text{ fb}^{-1}$ for the LHC running at $\sqrt{s}=14\text{ TeV}$.
We also give cross sections before cuts.\footnote{Only common generator-level cuts $p_T\geq 20\text{ GeV}$ and $\Delta R\equiv \sqrt{\Delta\phi^2 + \Delta\eta^2} \geq 0.4$ are imposed. Apart from that the values are calculated as described in Sec.~\ref{sec:Event_Generation}.}
Signal events are shown for two model points with $m_{t'}=800$ and $900\text{ GeV}$.
A breakdown of all considered background processes is given together with their sum (denoted as ``b.g.'').
A sufficiently large number of events which pass the $H_T\geq 1200\text{ GeV}$ cut are generated:
35587 events for $t'\bar{t'}$ (800~GeV), 40960 events for $t'\bar{t'}$ (900~GeV), 89046 events for $bb\bar{b}\bar{b}$, 76536 events for $t\bar{t}$, 27805 events for $t\bar{t}b\bar{b}$, and 12190 events for $t\bar{t}h$. 

As can be seen from Table~\ref{tab:result_nofevents}, the first two simple cuts ($H_T \geq 1200$~GeV and $\# b\geq4$) already drastically suppress the various backgrounds.
After top and Higgs tagging and subsequent reconstruction of the vectorlike top, there is a clear excess of signal events over background in both signal regions SR1 ($1t$+$2h$) and SR2 ($2t$+$1h$).
For the considered integrated luminosity, event numbers are small if we require both two reconstructed tops and Higgs bosons (SR3).
The number of signal events for 800~GeV turns out to be less than two with almost vanishing backgrounds $< 0.35$.
These numbers are too small to reconstruct a mass peak for the vectorlike top.

\begin{figure}[hbtp]
 \begin{center}
  \includegraphics[scale=.6]{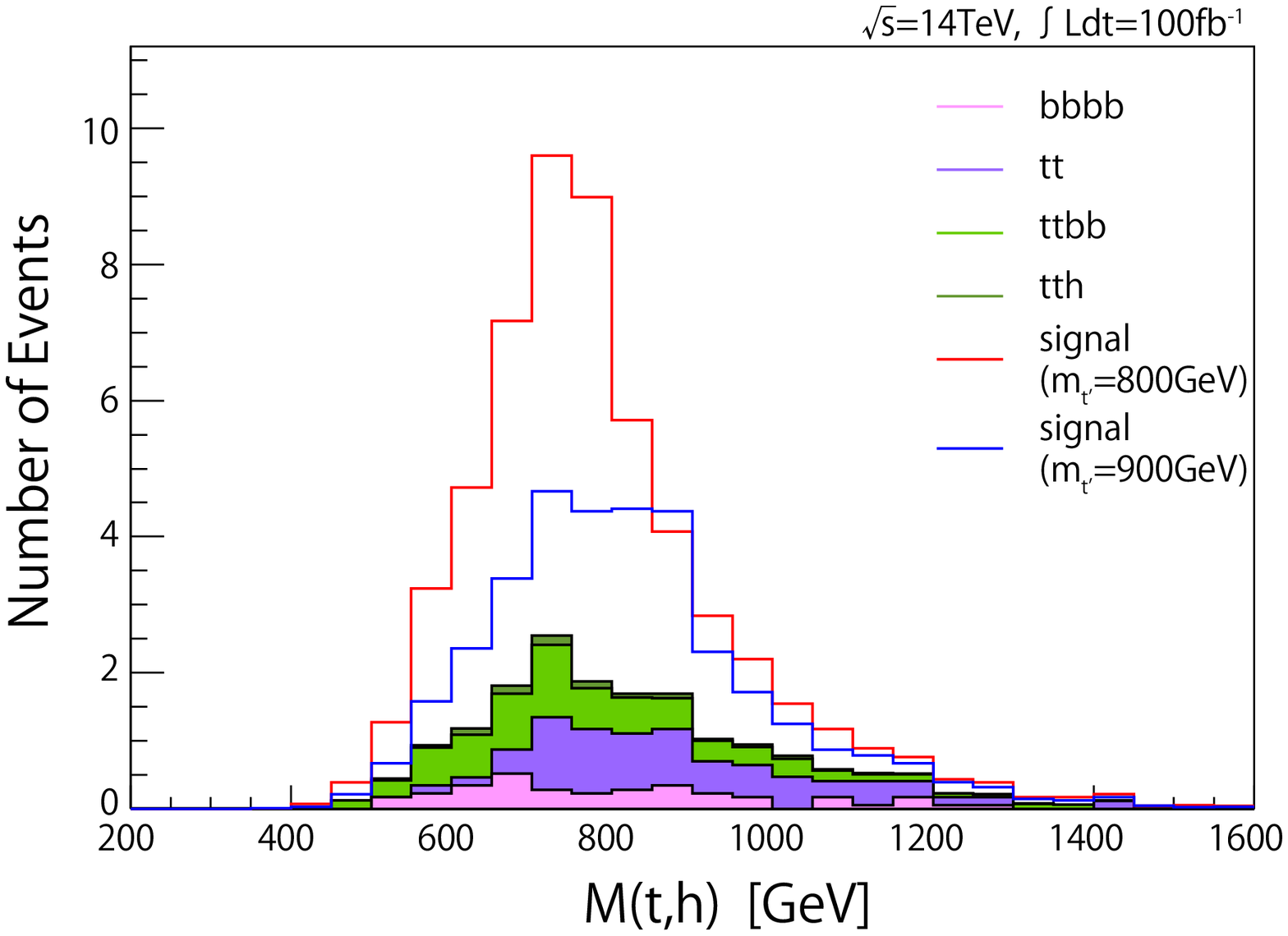}
  \\
  \includegraphics[scale=.6]{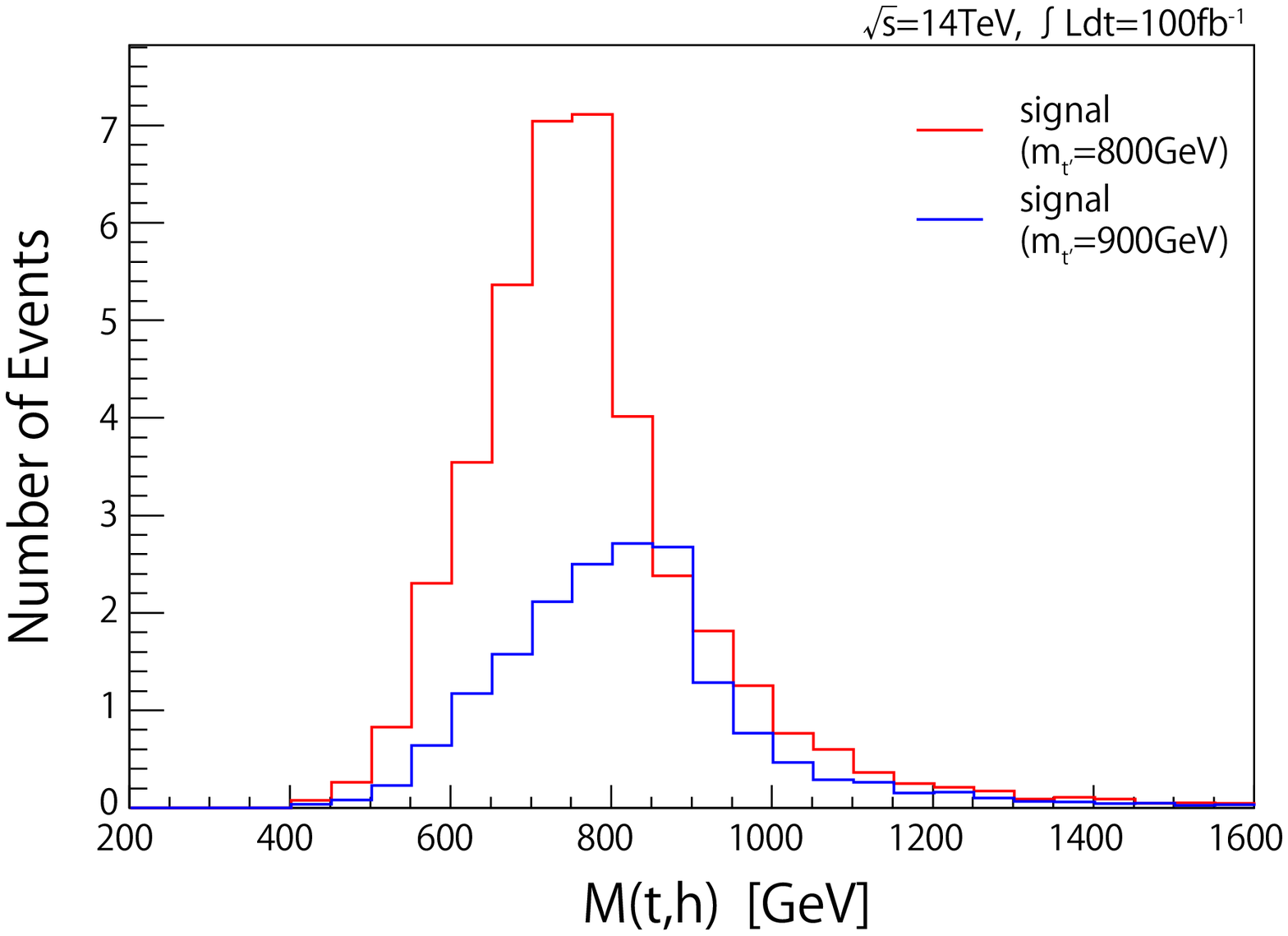}
 \end{center}
 \caption{(upper) The mass distribution of reconstructed vectorlike tops (detector level). The red and blue lines correspond to the signal for different masses of the vectorlike top, $m_{t'}=800\text{ GeV}$ and $m_{t'}=900\text{ GeV}$, respectively. The black line shows the sum of all relevant background processes; their breakdown is given by the filled curves. Event numbers are stacked. (lower) The mass distribution for signal events only  (detector level). }
 \label{fig:result_peak}
\end{figure}

For the combined signal region SR1+SR2, the reconstructed mass distribution is given in Fig.~\ref{fig:result_peak}.
The red line corresponds to the signal with mass 800~GeV and the blue line corresponds to the signal with mass 900~GeV. In the upper diagram  event numbers are stacked.
The black line shows the sum of all background processes; their breakdown is expressed by the filled curves.
The figure shows a clear mass peak of the vectorlike top. 
For the case of  $m_{t'} = 800\text{ GeV}$, the peak is around 700--800~GeV with a width of $\mathcal{O}(100)\text{ GeV}$ and experiences a steep drop just above the true mass.
Falsly-assigned $t$-$h$ pairs typically lead to an overestimation of the values of the reconstructed mass.
The lower cutoff in background events is dominated by the cut on scalar transverse momentum $H_T$.
The shape of the peak is also affected by the accuracy of reconstructed tops and Higgs bosons.
Tighter mass ranges in the tagging algorithms do lead to a sharper mass peak, but at the cost of decreasing event rates.

For signal events with a vectorlike top mass of 900~GeV the excess over background is smaller.
The reconstructed mass peak is lower and wider, but again experiences a sharp edge just above the true mass (see Fig.~\ref{fig:result_peak} lower panel).
Since the initial $H_T$ distribution is shifted to larger values by 200~GeV compared to $m_{t'}=800\text{ GeV}$ (see Fig.~\ref{fig:cut_ht_dist}), a stricter cut on scalar transverse momentum (e.g.~$H_T \geq 1400$~GeV) can improve the signal-to-background ratio.
As the event numbers also drop, this cut should be considered only for larger integrated luminosities.

\section{Summary and outlook}
\label{sec:summary_outlook}

We investigated fully hadronic final states to search for pair-produced vectorlike top partners at the LHC.
Imposing an $H_T$ cut, multibottom cut, and using top~/~Higgs taggers we can suppress the background processes and reconstruct the vectorlike top.
For this reconstruction we proposed an algorithm to determine the $t$-$h$ combination based on a massive pair hypothesis.
We note that our analysis procedure is kept general and in particular not tailored to any model parameter or mass scale except for the initial $H_T$ cut.
It was found that the vectorlike top can be reconstructed with good quality and signal-to-background ratio if $\text{BR}(t'\rightarrow t h)$ is large.
 
Although we considered fully hadronic final states, our algorithm can be applied to events with semileptonically decaying vectorlike top quarks as well.
 
In this paper we assumed $\text{BR}(t'\rightarrow t h)=1$.
A complete analysis should also cover the cases of generic branching fractions to other possible final states such as $t'\rightarrow b W$ and $t'\rightarrow t Z$.
The former decay leads to quite distinct final states, but the latter one can lead to similar event topologies as the $th\bar{t}h$ final states, which affects the result of our analysis.  
Even without considering mistags, the decay chain $t'\bar{t'} \rightarrow (th)(\bar{t}Z)$ can give a contribution to SR2.
This will lead to a wrong assumption on the untagged particle's mass when determining the $t$-$h$ combination from Eqs.~(\ref{eq:mass_1}) and (\ref{eq:mass_2}).
On the other hand, due to the loose mass constraints employed in the Higgs boson tagging algorithm, misidentification of $Z$ as $h$ leads to a broadened mass peak for the vectorlike top.
Its shape may act as a handle on determining the correct branching fractions, in conjunction with event counts.
As we only give a proof of concept here, a detailed analysis is left for future studies.

\section*{Acknowledgment}
The authors are grateful to Michael Spannowsky and Yoshitaro Takaesu for helpful discussions.
This work was supported by the Program for Leading Graduate Schools (M.S.), JSPS KAKENHI Grant No. 23740172 (M.E.)  and No. 26800123 (K.H.), and also by the World Premier International Center Initiative (WPI Program), MEXT, Japan.

\providecommand{\href}[2]{#2}\begingroup\raggedright\endgroup

\end{document}